\documentclass[final]{aipproc}
\layoutstyle{6x9}
\usepackage{graphicx}
\begin{document} 
\title{The constraint on the spin dependent structure function $g_1$ at low $Q^2$
through the sum rule corresponding to the moment at $n=0$}
\author{Susumu Koretune}{
address={Department of Physics, Shimane University,Matsue,Shimane,690-8504,Japan}}
\begin{abstract} 
 The sum rules for the spin dependent structure function $g_1^{ab}$ in the null-plane
formalism corresponding to the moment at $n=0$ has been transformed
to the sum rule which relates the $g_1^{ab}$ with the cross section of the
isovector photon or the real photon. Based on these sum rules, we argue 
that there is a deep connection among the elastic,
the resonance, and the non-resonant contributions,
and that it explains why the sign of the generalized
Gerasimov-Drell-Hearn sum changes at very small $Q^2$. 
\end{abstract}

\maketitle
\section{Introduction}
The fact that the sign of the Gerasimov-Drell-Hearn(GDH)
sum rule\cite{Drell,Ger} and that of the Ellis-Jaffe sum
rule\cite{Ellis} was different had motivated
the study of these sum rules and the spin structure functions $g_1$ and $g_2$
at low $Q^2$. 
Experimentally, the sign of the $ \Gamma^P(Q^2)$ defined as 
\begin{equation}
 I^p(Q^2)=\frac{2m_p^2}{Q^2}\Gamma^P(Q^2),
\end{equation}
\begin{equation}
 \Gamma^P(Q^2) = \int_0^1dx g_1^p(x,Q^2),
\end{equation}
where the $I_P(0)$ is known to be negative through the GDH sum rule
was studied at CLAS\cite{Fatemi}, and was shown that it changed a sign in the
very small $Q^2$ region. Further the large negative contribution in
this region was shown to become
small quickly as we go to the $Q^2$ near 1 (GeV/c)$^2$.
In this talk we show that this
rapid change is tightly connected to the rapid change of the elastic
contribution in this region\cite{kore1,kore2}.
\section{The sum rule in the isovector reaction}
The sum rule for the $g_1^{ab}$ is derived from 
the current commutation relation on the null-plane for the good-bad
component $ [J_a^{+}(x),J_b^i(0)]|_{x^+=0}$\cite{DJT}, and given as
\begin{equation}
 \int_{0}^{1}\frac{dx}{x}g_1^{[ab]}(x,Q^2)=-\frac{1}{16} f_{abc}\int_{-\infty}^{\infty}
d\alpha [A_c^5(\alpha ,0)+\alpha \bar{A}_c^5(\alpha ,0)],
\end{equation}
where $A_c^5(\alpha ,0)$ and $\bar{A}_c^5(\alpha ,0)$ is the matrix
element of the bilocal current,
and $g_1^{ab}$ is defined as
\begin{eqnarray}
W_{\mu\nu}^{ab}|_{spin}&=&\frac{1}{4\pi}\int d^4x\exp (iqx)
<p,s|J_{\mu}^a(x)\cdot J_{\nu}^b(0)|p,s>_c |_{spin}\\
&=&i\epsilon_{\mu\nu\lambda\sigma}q^{\lambda}s^{\sigma}G_1^{ab}
+i\epsilon_{\mu\nu\lambda\sigma}q^{\lambda}(\nu s^{\sigma}-
q\cdot sp^{\sigma})G_2^{ab},
\end{eqnarray}
with $g_1^{ab}=\nu G_1^{ab}$ and $g_2^{ab}=\nu^2G_2^{ab}$.
Since the right hand side of Eq.(3) is $Q^2$ independent, we obtain for the
anti-symmetric combination with respect to $a,b$
\begin{equation}
 \int_{0}^{1}\frac{dx}{x}g_1^{[ab]}(x,Q^2)=\int_{0}^{1}\frac{dx}{x}g_1^{[ab]}(x,Q^2_0).
\end{equation}
Now the Regge theory predicts as 
$g_1^{[ab]}\sim \beta x^{-\alpha(0)}$ with $\alpha(0) \leq 0$,
and hence the sum rule is convergent.
However, the perturbative behavior like the DGLAP is
divergent. The double logarithmic $(log (1/x))^2$ resummation 
gives more singular behavior than the Regge
theory\cite{bade} and the sum rule is also divergent.
Though, whether the sum rule diverges or not can not be judged
rigorously by these discussions, it is desirable to discuss the 
regularization of the sum rule and gives it a physical meaning even 
when the sum rule is divergent. Now, the regularization of the 
divergent sum rule has been known to be done by the analytical
continuation from the nonforward direction\cite{analy}.
We first derive the finite sum rule in the small but sufficiently large $|t|$ region
by assuming the moving pole or cut. Then we subtract the singular pieces
which we meet as we go to the smaller $|t|$ from both hand sides of the
sum rule by obtaining the condition for the coefficient of the singular
piece. After taking out all singular pieces we take the limit $|t|\to
0$. Because of the kinematical structure in the course to derive the sum
rule, we can mimic this procedure in the forward direction by introducing the 
parameter which reflects the $t$ in the non-forward direction.
The sum rule obtained in this way can be transformed to the form
where the high energy behavior from both hand sides of the sum rule
is subtracted away.  Practically, if the cancellation at high energy is
effective, since the condition is needed only in the high energy limit, 
we consider that the sum rule holds irrespective of the condition. In this
way, we subtract the high energy behavior from both hand sides of
Eq.(6).

Now we take $Q_0^2=0$ and use the relation
\begin{equation}
 G_1^{ab}(\nu ,0)=-\frac{1}{8\pi^2\alpha_{em}}\{\sigma_{3/2}^{ab}(\nu ) - \sigma_{1/2}^{ab}(\nu )\}
=-\frac{1}{8\pi^2\alpha_{em}}\Delta\sigma^{ab}(\nu ).
\end{equation}
Then we define $\nu_{c}^{Q} =m_pE_Q$ where
$E_Q=E_c+Q^2/2m_p$ and $E$ is a energy of the real(virtual) photon in the laboratory frame. 
By setting $a=(1+i2)/\sqrt{2}, b=a^{\dagger}$, and using the same
method as in the Cabibbo-Radicati sum rule\cite{cab}, we obtain the sum rule which relates the
$g_1$ and the cross section of the isovector photon
by separating out the elastic contribution as
\begin{eqnarray}
\lefteqn{\int_{x_c(Q)}^{1}\frac{dx}{x}[2g_1^{1/2}(x,Q^2)-g_1^{3/2}(x,Q^2)]}&&\\\nonumber
&=&B(Q^2)- \frac{m_p}{8\pi^2\alpha_{em}}\int_{E_0}^{E_c}dE[2\Delta\sigma^{1/2}-\Delta\sigma^{3/2}]+K(E_c,Q^2),
\end{eqnarray}
where $x_c(Q^2)=\frac{Q^2}{2\nu_{c}^{Q}}$ and
\begin{equation}
B(Q^2)=\frac{1}{4}\{(\mu_p-\mu_n)-\frac{1}{1+Q^2/4m_p^2}G_{M}^{+}(Q^2)
[G_{E}^{+}(Q^2)+\frac{Q^2}{4m_p^2}G_{M}^{+}(Q^2)]\},
\end{equation}
\begin{equation}
G_E^+(Q^2)=G_E^p(Q^2)-G_E^n(Q^2), \qquad
G_M^+(Q^2)=G_M^p(Q^2)-G_M^n(Q^2) ,
\end{equation}
\begin{equation}
K(E_c,Q^2) =
 -\int_{E_Q}^{\infty}\frac{dE}{E}[2g_1^{1/2}(x,Q^2)-g_1^{3/2}(x,Q^2)] 
 - \frac{m_p}{8\pi^2\alpha_{em}}\int_{E_c}^{\infty}dE[2\Delta\sigma^{1/2}-\Delta\sigma^{3/2}] ,
\end{equation}
with $g_1^I$ and $\Delta\sigma^I$ being the quantities in the reaction\\
$ ``{\rm isovector\; photon\; +\; proton\; \longrightarrow state\; with\; isospin\; I }.``$\\
It should be noted that, because of the reguralization of the sum rule,
we are allowed to take the integral over $E$ in $K(E_c,Q^2)$ after the subtraction.
Since $K(E_c,Q^2)$ is expected to be small, we have the relation 
among the elastic, the resonance, and the non-resonant contributions.
\section{The sum rule in the elctroproduction}
The current anticommutation relation on the null-plane was derived using
the DGS representation\cite{DGS,kore80,kore84,kore93}. It should be noted that
this relation is not the operator relation ant that it holds only
for the matrix element between the stable
one particle hadronic state.
The sum rule in this case takes the form
\begin{equation}
 \int_0^1\frac{dx}{x}g_1^{ab}(x,Q^2)=-\frac{1}{8\pi}d_{abc}\int_{-\infty}^{\infty}d\alpha
\ln |\alpha |\{S_c^5(\alpha ,0) + \alpha \bar{S}_c^5(\alpha ,0))\},
\end{equation}
\begin{equation}
  \int_0^1\frac{dx}{x}g_1^{p}(x,Q^2)= \int_0^1\frac{dx}{x}g_1^{p}(x,Q_0^2),
\end{equation}
where $S_c^5(\alpha ,0)$ and $\bar{S}_c^5(\alpha ,0)$ is defined similarly
as $A_c^5(\alpha ,0)$ and $\bar{A}_c^5(\alpha ,0)$. 
By separating out the Born term, and using the same method as in the
current commutator case we obtain
\begin{equation}
\int_{x_c(Q^2)}^1\frac{dx}{x}g_1^{p}(x,Q^2)= B(Q^2)
-\frac{m_p}{8\pi^2\alpha_{em}}\int_{E_0}^{E_c}dE\{
\sigma_{3/2}^{\gamma p} - \sigma_{1/2}^{\gamma p}\}+K(E_c,Q^2),
\end{equation}
where
\begin{equation}
B(Q^2)=\frac{1}{2}\{\mu_p -\frac{1}{1+Q^2/4m_p^2}G_{M}^{p}(Q^2)
[G_{E}^{p}(Q^2)+\frac{Q^2}{4m_p^2}G_{M}^{p}(Q^2)]\},
\end{equation}
\begin{equation}
K(E_c,Q^2)=\frac{m_p}{8\pi^2\alpha_{em}}\int_{E_c}^{\infty}dE\{
\sigma_{1/2}^{\gamma p} - \sigma_{3/2}^{\gamma p}\}
-\int_{E_c^Q}^{\infty}\frac{dE}{E}g_1^{p}(x,Q^2).
\end{equation}
Similar sum rule can be derived in case of the neutron target. However
it should be noted that, in this
case, the corresponding term of the $\mu_p$ in the Born term 
becomes $0\times\mu_n=0$ since $G_E^n(0)=0$.\\

Now, using parameter in \cite{sim}, we find that $K(2,Q^2)$ below $Q^2=0.5$(GeV/c)$^2$ is
very small. Further, by the experimental data from GDH
collaboration\cite{dutz}, we find
\begin{equation}
\frac{m_p}{8\pi^2\alpha_{em}}\int_{E_0}^{2}dE\{
\sigma_{3/2}^{\gamma p} - \sigma_{1/2}^{\gamma p}\}
\sim 0.45,
\end{equation}
with the possible error about 20\%.
We use the standard dipole fit for the Sachs form factor
and find that, in the small $Q^2$ region near $Q^2\sim 0.1$(GeV/c)$^2$, 
the integral $\displaystyle{\int_{x_c(Q^2)}^1\frac{dx}{x}g_1^{p}(x,Q^2)}$
changes a sign. The integral is effectively cut off by the same $W^2=(p+q)^2$ both in the real
and the virtual photon. This means that the
same resonances contribute to the integral
$\displaystyle{\int_{x_c(Q^2)}^1\frac{dx}{x}g_1^{p}(x,Q^2)}$ 
both in the real and the virtual photon and hence this
integral measures the change of the resonances and the background 
in the very small $Q^2$ region. 
Thus its change is tightly related to the sign change of the GDH sum in this region.
\section{Conclusion}
We find, in the small $Q^2$ region near $Q^2\sim 0.1$(GeV/c)$^2$,
that the integral $\displaystyle{\int_{x_c(Q^2)}^1\frac{dx}{x}g_1^{p}(x,Q^2)}$
becomes zero and that it
changes a sign from the negative to the positive. 
This behavior is caused by the 
rapid change of the resonances together with the continuum to compensate
the rapid change of the elastic to satisfy the sum rule. 
It is this rapid change of the resonances which gives 
the sign change of the GDH sum. Hence we see why
it occurs in the very small $Q^2$ region.

\end{document}